**Polariton Induced Enhanced Emission from an Organic Dye under Strong Coupling Regime**

*Dario Ballarini\*, Milena De Giorgi, Salvatore Gambino, Giovanni Lerario, Marco Mazzeo, Armando Genco, Gianluca Accorsi, Carlo Giansante, Silvia Colella, Stefania D'Agostino, Paolo Cazzato, Daniele Sanvitto\*, Giuseppe Gigli*

D. Ballarini, M. De Giorgi, S. Gambino, M. Mazzeo, A. Genco, G. Accorsi, C. Giansante, S. Colella, P. Cazzato, D. Sanvitto, G. Gigli
NNL, Istituto Nanoscienze - CNR, Via Arnesano, 73100 Lecce, Italy
E-mail: dario.ballarini@nano.cnr.it, daniele.sanvitto@cnr.nano.it
D. Ballarini, M. De Giorgi, S. Gambino, G. Lerario, C. Giansante, S. D'Agostino, P. Cazzato, D. Sanvitto, G. Gigli
CBN, Istituto Italiano Tecnologia, Via Barsanti, 73010 Lecce, Italy
M. Mazzeo, G. Gigli
Dipartimento di Matematica e Fisica "Ennio De Giorgi", Università del Salento, Via Arnesano, 73100 Lecce, Italy

Exciton-polaritons in semiconductors are quasi-particles which have recently shown the capability to undergo phase transition into a coherent hybrid state of light and matter. The observation of such quasi-particles in organic microcavities has attracted increasing attention for their characteristic of reaching condensation at room temperature. In this work we demonstrate that the emission properties of organic polaritons do not depend on the overlap between the absorption and emission states of the molecule and that the emission dynamics are modified in the strong coupling regime, showing a significant enhancement of the photoluminescence intensity as compared to the bare dye. This paves the way to the investigation of molecules with large absorption coefficients but poor emission efficiencies for the realization of polariton condensates and organic electrically injected lasers by exploiting strong exciton-photon coupling regimes.

# 1. Introduction

The emission properties of an optically active material in a Fabry-Perot resonator are strongly modified by the coupling of the excitonic transition with the confined electromagnetic mode, both in the weak and in the strong coupling regimes. While the weak coupling involves directly the emission of a photon and the final density of states of the electromagnetic field, the strong coupling expresses a deeper interaction between light and matter, which results in the appearance of a new set of eigenmodes, namely the upper (UPB) and lower (LPB) polariton branches.[1,2] In the polariton picture, energy is coherently exchanged between excitons and photons through periodic oscillations (also called Rabi oscillation), which can be qualitatively described by the coherent emission and re-absorption of a photon by the material dipole. In inorganic semiconductors (usually quantum wells) the absorption and emission occur almost at the same energy, whereas in most organic molecules a conformational rearrangement and energy relaxation follow the absorption of light, which can lead to a substantial shift between the emission and the absorption peaks. In order to reach the best approximation to a perfect 2-level system, J-aggregates have been widely used for their sharp absorption peak and small Stokes shift.[3-9] However, the observation of polariton resonances using transmission and emission measurements in molecules with a large energy shift between absorption and emission[10-13] indicates that only the absorption configuration of the molecule is strongly coupled with the cavity mode. This coherent absorption-emission loop is an actual new state of the system, which can decay radiatively. If the competition with non-radiative channels is won by radiative polariton decay, the emission efficiency could, in principle, be enhanced. This has been already discussed in the early work of Reference [14] for the case of a porphyrin dye, but never observed so far.[6, 15-22]

Here we demonstrate an enhancement of the emission intensity of more than one order of magnitude by using a cavity made by silver thin mirrors with a quite low finesse (Q=20) and a squaraine dye molecule, whose absorption transition is strongly coupled with the cavity mode. Time resolved measurements show that the emission dynamics are modified under strong coupling regime, allowing the circumvention of non-radiative channels through the relaxation into radiative polariton modes.

## 2. Results

The experiments are performed on a squaraine dye (**Figure 1**a) co-evaporated in a NPB matrix (shown in Figure S1, Supporting Information) in a 10-to-90 mass ratio (henceforth referred to as SQ). The absorption and photoluminescence (PL) spectra of a 130~nm-thick SQ film are shown in Figure 1b. The SQ presents an absorption resonance at 1.85 eV and an emission peak at 1.80 eV, showing an energy shift between absorption and emission of about 50 meV.

When the SQ thin-film is placed between two evaporated silver mirrors of 35 nm thickness (henceforth referred to as SQMC), a cavity mode, whose energy is on resonance with that of the SQ absorption peak, brings the system into the strong coupling regime. The UPB and LPB are clearly observable through the transmittance measurements shown in Figure 1c, where the typical anticrossing behavior is reproduced as a function of the in-plane momentum $k = \frac{\omega}{c}\sin(\theta)$, obtained by changing the angle of incidence θ.

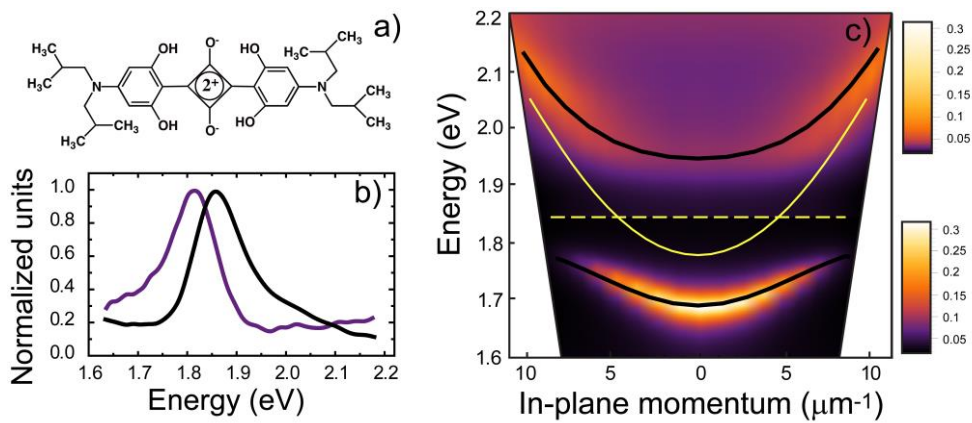

**Figure 1.**

a) Molecular structure of the squaraine dye b) PL (purple line) and absorption spectra (black line) of the co-evaporated squaraine (10%) in NPB matrix (90%) thin-film (SQ). c) In the background, angle-resolved transmittance spectra show the dispersion of the UPB (in logarithmic color scale) and LPB (in linear color scale) as a function of the in-plane momenta for a cavity thickness of ≈140 nm (TE polarization). Experimental data are fitted by transfer matrix calculations (black lines for the UPB and LPB), while the bare cavity and exciton modes are shown as yellow and dashed-yellow lines, respectively.

Experimental polariton dispersions, shown as background in Figure 1c, are well reproduced by the SQMC transmittance obtained by transfer matrix calculation, as shown in Figure 1c by black lines (UPB and LPB) and yellow line (cavity mode), while the molecular absorption is indicated by the horizontal dashed line.

The emission properties of SQMC and SQ samples have been investigated by using a continuous wave laser tuned at 3.06 eV, well above the UPB and where the reflection of the silver mirror is largely reduced (R=65%). Photons are absorbed at this energy by the NPB matrix (O.D.=0.2 at 3.06 eV), while the population of the excited states of the squaraine is partially allowed by energy transfer (Figure S1, Supporting Information). In **Figure 2**a, the transmission and emission energies of the SQMC and that of the SQ reference sample (same SQMC but without the top mirror) are compared for different film thicknesses.

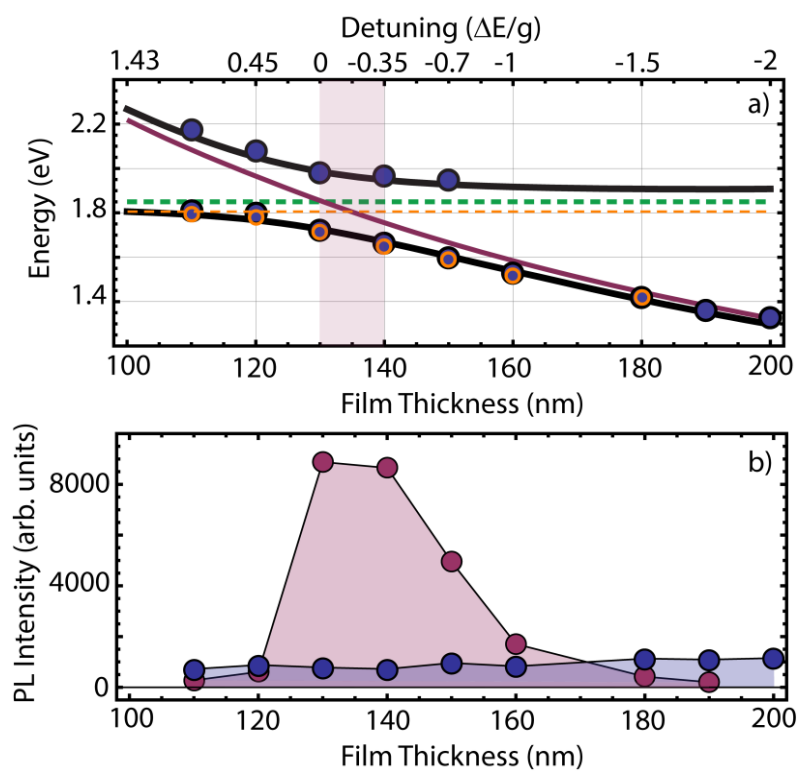

**Figure 2.**

a) The energies of the LPB emission (open orange circles) in the SQMC sample are compared with the energies of LPB and UPB transmission (full dots) as a function of the film thickness at k=0. The absorption and emission of the reference samples are indicated by dashed horizontal lines in green and orange colors, respectively. The LPB and UPB energies as obtained by transfer matrix calculations are represented by black lines, while the cavity mode is shown by purple line which intersect the absorption resonance of the molecule for a film thickness of 130 nm. The filled area between 130 nm and 140 nm corresponds to the region of maximum emission enhancement. b) The PL intensity of the SQMC (purple dots) and SQ reference (blue dots), as a function of the sample thickness, shows a strong enhancement between 130 nm and 140 nm.

The transmission energies of the UPB and LPB are shown as full blue dots, while the results of transfer matrix calculation are shown as black line (UPB and LPB) and red line (cavity mode). By changing the cavity thickness from 100 nm to 200 nm, the detuning of the cavity mode with respect to the absorption transition of the dye, defined as $\Delta E = E_{cav} - E_{ex}$, spans from positive to negative values, with the zero-detuned SQMC at 130 nm. The corresponding Rabi splitting, defined as the minimum energy splitting between the UPB and LPB, is g=260 meV. Due to this large coupling strength, typical of organic microcavities, the value $\frac{\Delta E}{g}$ is significantly altered only if the cavity mode is spanned on an energy range of hundreds of meV, resulting in a nonlinear dependence of $\frac{\Delta E}{g}$ on the cavity thickness as shown in Figure 2a. The PL energies of the SQMC (empty orange circles in Figure 2a) follow the transmission resonances of the LPB, confirming that both emission and transmission are under strong coupling regime. PL and absorption energies of the SQ reference samples are instead independent on the film thickness, and are indicated by dashed orange and green lines, respectively.

The maximum emission intensity occurs around the k=0 directions of the LPB for all the detunings considered in Figure 2. The emission intensities of the SQMC and those of the reference SQ, at k=0, are compared in Figure 2b as a function of the film thickness. Remarkably, the PL intensity of the LPB shows a clear maximum around 130-140 nm, corresponding to the shadowed region in Figure 2a ( $\Delta E/g = \{-0.35, 0\}$), whereas the PL intensity of the reference SQ film is almost constant for the film thicknesses considered. We note also that the enhancement occurs only when the energy of the LPB is lower than that of the dye emission peak, $E_{LPB} < 1.85$ eV, with a sharp increase in correspondence of the 130 nm thick sample. For smaller cavity thicknesses, where the LPB is resonant with the emission energy of the dye (110 nm and 120 nm in Figure 2), the SQMC emission intensities are lower than that of the reference samples. For larger cavity thicknesses ($\geq$ 150 nm), the intensity of the polariton emission smoothly decreases, suggesting a reduced scattering probability towards the bottom of the LPB.[23, 24, 25]

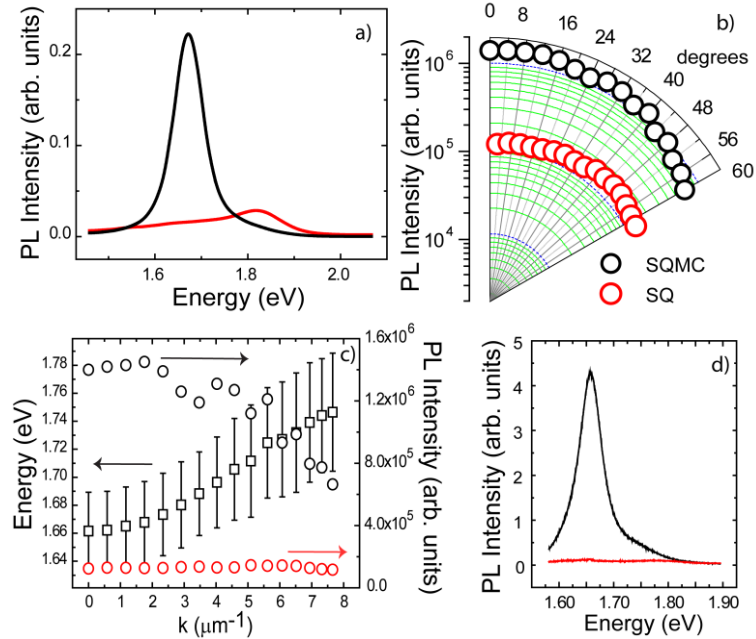

**Figure 3.**

a) Emission intensity of the SQMC and SQ reference at k=0 for a 140 nm thick sample. The LPB emission (black line) is 60 nm red-shifted compared to the emission energy of the SQ reference (red line) and shows an increase of ten times of the integrated PL intensity. b) Angle-resolved emission intensities, taken at the energies of the LPB for the SQMC (black dots) and for the SQ reference (red dots), with a film thickness of 140nm. c) The energies of the LPB are plotted as a function of the in-plane momentum k, with the FWHM indicated by vertical bars (left axes). The corresponding PL intensities of SQMC (black circles) and reference (red circles) are shown on the right axes. d) PL intensities of SQMC (black line) and reference SQ (red line) obtained from emission around k=0 (acceptance angle: ±14°) when the excitation energy is tuned on resonance with the UPB (1.96 eV).

As shown in the emission spectra of **Figure 3**a, the PL intensity of SQMC along the normal to the sample surface (acceptance angle $\pm14°$) is more than 10 times higher if compared to the reference SQ film of the same thickness (140 nm). However, in analyzing these results, care must be taken to distinguish the effects of the strong coupling on the total PL intensity from those induced by a redistribution of the angular intensity around k=0.[26-30]

In Figure 3b, the PL intensities at the energy of the LPB are plotted as a function of the emission angle, showing that the integrated intensities in the range from 0° to 60° still give an overall increase of one order of magnitude in the case of SQMC, therefore excluding that the enhancement can be ascribed only to the modification of the angular emission pattern.[31, 32, 33] For comparison, the intensity redistribution expected in the weak coupling regime is shown in Figure S3 of the Supporting Information. In Figure 3c, the PL intensities as a function of the in-plane momentum are shown (black and red circles for SQMC and SQ, respectively) with the corresponding LPB energy dispersion (black squares), whose full width at half maximum (FWHM) is indicated by vertical bars. At higher in-plane momenta, the intensity is reduced to half of the intensity along the cavity axis (k=0), while the exciton weight of the LPB is doubled from 0.35 at k=0 to 0.7 at k=7 $\mu m^{-1}$. In order to exclude any hidden losses for the emission of the SQ sample, the sphere-integrated photoluminescence quantum yield (PLQY) has been measured for the cavity and reference samples.[34] Under non-resonant excitation, the total yield is determined by the combination of the rate of energy transfer from the host to the acceptor and the intrinsic PLQY of the dye, resulting in a very low total efficiency for which we can only measure the instrument background, giving an upper limit of $\Phi_{SQ}<0.01\%$. When the film is placed inside the cavity, the PLQY under the same conditions increases to the measured value of $\Phi_{SQMC}=(0.03\pm0.01)\%$ for the SQMC of 140 nm, endorsing a better extraction of light in the system under strong coupling regime. However, if the SQMC is resonantly excited at the energy of the UPB (1.96 eV), the difference between the PL intensity of the SQMC and reference is much larger, as can be seen from the spectra of Figure 3d.

While an absolute comparison of the emission efficiencies cannot be done under these excitation conditions due to the presence of the mirrors, it is worth noting that the PLQY of the SQMC shows almost an order of magnitude enhancement with respect to the non-resonant pumping configuration, rising up to $\Phi_{SQMC}$=0.2% and confirming that polaritons can be resonantly injected into the UPB and efficiently transferred to the LPB via scattering from the exciton reservoir.[6, 7]

As shown in Figure S4 of the Supporting Information, the redistribution of the radiation pattern induced by the polariton resonances cannot explain alone the observed results without specifically address the new relaxation dynamics of the strongly coupled system, which will be addressed in the following of this work.

To exclude that the enhancement is originated by a change in the rate of energy transfer from the NPB to the squaraine due to microcavity effects,[35, 36] time resolved PL measurements (TRPL) have been performed to compare the decay time of the NPB emission (detection at 2.8 eV) in the case of the reference sample and the SQMC. As shown in Figure S5 of the Supporting Information, there is not any appreciable difference between the two curves, which can be fitted by the same bi-exponential function with times $t_1$=30 ps and $t_2$=170 ps, showing that the energy transfer is not affected by presence of the cavity.

In **Figure 4**a, TRPL are shown for a 140 nm-thick reference SQ at 1.8 eV (green line) and SQMC with different thicknesses: 130 nm (red line), 140 nm (blue line), 150 nm (orange line). The decay time of the PL intensity in SQ samples does not change with the thickness of the film (not shown) and it is much shorter than the radiative lifetime of the dye (of the order of 2 ns), indicating that it is limited by faster non-radiative channels. In our case, the decay of the SQ PL shown in Figure 4a is mainly given by the relaxation dynamics, while the non-radiative processes (faster than 15 ps) are responsible for the fast PL rise time of SQ films. Both the PL rise time and decay time of SQMC samples are instead slower than those of the reference sample. In Figure 4b, the dynamics of the first 40 ps of the TRPL of SQ (green line) and SQMC (130nm, 140nm, 150nm with red, blue and orange line, respectively) are shown.

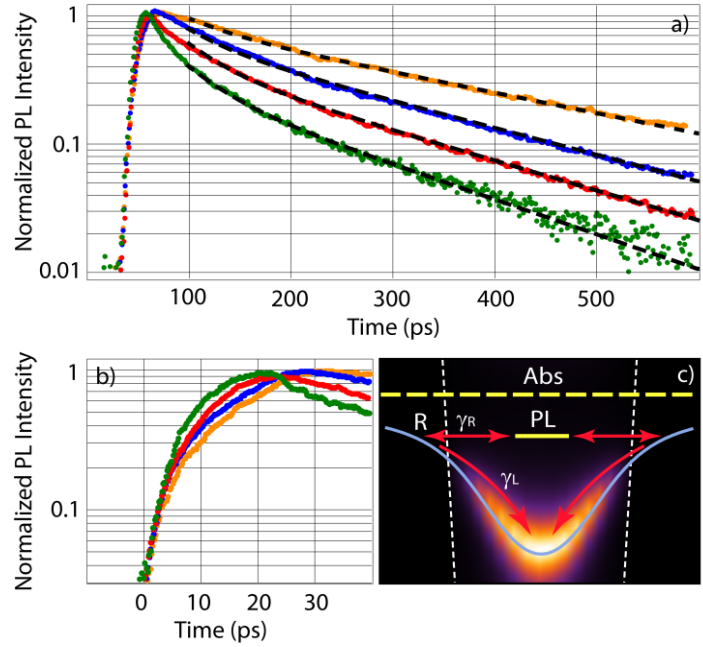

**Figure 4.**

a) TRPL of the reference sample (green line), SQMC with 130 nm thickness (red line), 140 nm thickness (blue line) and 150 nm thickness (orange line). Black lines are the results of the rate equations (1-3) with the following parameters: $\gamma_0=1/2000$ ps$^{-1}$, $\gamma_{NR}=1/10$ ps$^{-1}$, $\gamma_{Ph}=1/0.05$ ps$^{-1}$ for all the curves, while $\gamma_R=0$, $\gamma_R=1/15$ ps$^{-1}$, $\gamma_R=1/17$ ps$^{-1}$, $\gamma_R=1/22$ ps$^{-1}$ and $\gamma_L=0$, $\gamma_L=1/31$ ps$^{-1}$, $\gamma_L=1/47$ ps$^{-1}$, $\gamma_L=1/150$ ps$^{-1}$ for the SQ reference, SQMC of 130 nm, 140 nm and 150 nm thickness, respectively. P is obtained from the bi-exponential decay of NPB with $t_1=30$ ps and $t_2=170$ ps for all curves, while $P_{sq}$ from the SQ decay at 1.7 eV ($t_{sq}=250$ ps) and $\gamma_R$ ($\gamma_{NR}$) from the rise times of the LPB (SQ) shown in panel b). b) TRPL of the first 40 ps evidencing the slowing of the rise time when increasing the negative detuning: SQ (green line) and SQMC (130nm, 140nm, 150nm with red, blue and orange line, respectively), with rising time going from 15 ps to 25 ps for thicker cavities. c) Simplified model of the relaxation from uncoupled excitons (level "PL") into the bottom of the LPB. The dispersion of the LPB for the 140 nm thick SQMC is reported in the background.

Even if direct scattering from uncoupled molecules to the bottom of the LPB is an allowed process, the slower dynamics of the SQMC, as compared to those of SQ, indicates that the reservoir cannot be identified with the PL state of the molecule: due to the fast radiative decay of polaritons, the population of the LPB follows quickly the population of its feeding state. This would result to a faster rather than slower decay dynamics. The increase in the decay time is instead consistent with the presence of additional states, which acts as an effective reservoir for polaritons at the bottom of the LPB. A qualitative understanding of the relaxation dynamics can be gathered by the simplified model sketched in Figure 4c, where the dashed yellow line corresponds to the absorption transition of the molecule, which is split in the SQMC samples into the UPB and LPB. Under non-resonant excitation, molecules relax into the lowest energetic configuration, which is uncoupled to light (PL level in Figure 4c), or coupled with light at high k-vectors (reservoir R in Figure 4c). The bottom of the lower polariton branch (LPB) can then be populated by two main channels: phonon-assisted scattering and optical pumping.[9, 24-25, 37-40] While the first mechanism can effectively contribute to an enhancement of the PL, the second one does not change significantly the total amount of emitted light. Although a complete model would include also channels for direct scattering from PL to LPB and for relaxation of excitons into the reservoir, given their negligible contribution to the dynamics, the traces shown in Figure 4a can be easily reproduced by a simpler three levels rate equations where the intermediate state, called reservoir in Figure 4c, would contribute to the filling of polaritons at the bottom of the LPB. This is done in analogy to other room temperature polariton systems (ZnO, GaN) and accordingly to recent observations of a polariton bottleneck also in organic microcavity.[1, 24, 41, 42] Being $n_{SQ}$, $n_R$ and $n_{LPB}$ the population of excitons, the reservoir and the bottom of the LPB, respectively, the rate equations can be written as:

$$\frac{\partial n_{SQ}}{\partial t} = -(\gamma_0 + \gamma_{NR} + \gamma_R)n_{SQ} + P + \gamma_R n_R \quad (1)$$

$$\frac{\partial n_R}{\partial t} = -(\gamma_R + \gamma_L + \gamma_{NR})n_R + \gamma_R n_{SQ} \quad (2)$$

$$\frac{\partial n_{LPB}}{\partial t} = -\gamma_{Ph} n_{LPB} + \gamma_L n_R \quad (3)$$

Here, P represents the relaxation from higher-energy excited-states, $\gamma_0$ and $\gamma_{NR}$ are the radiative and non-radiative decay rates of the bare molecule, $\gamma_R$ the interaction rate with the exciton-like reservoir, $\gamma_L$ the scattering rate towards the bottom of the LPB and $\gamma_{Ph}$ is the radiative decay rate of the LPB. In addition, the weak SQ emission at the lower energy tail of the PL peak (see Figure 1a) can populate the LPB below the bottleneck region by optical pumping. Indeed, the slower PL decay of SQ at longer wavelengths contributes to the SQMC emission when the scattering from the reservoir becomes inefficient and it is taken into account by adding an additional pumping term $P_{sq}$ to Equation (3) with relative weights of 0.003 P, 0.007 P and 0.01 P for 130 nm, 140 nm and 150 nm SQMC, respectively. While this contribution does not affect significantly the enhancement of the total PL intensity, it should be taken into account for better fitting the SQMC decay at negative detunings (thickness of the cavity ≥140 nm).

The black lines in Figure 4a are the results of the rate equations for the SQ reference and SQMC of 130 nm, 140 nm and 150 nm with the parameters given in the figure caption. The scattering rate from reservoir into the LPB, $\gamma_L$, is left as a free parameter and decreases at negative detunings, as expected for phonon-mediated processes. Considering that the radiative decay rate of the dye, $\gamma_0$, is much smaller than the corresponding non-radiative decay rates $\gamma_{NR}$, and that the radiative decay rate of the LPB, $\gamma_{Ph}$, is very large, being basically given by the photon lifetime in the cavity (few femtoseconds in our low-Q cavity, which gives $\gamma_{Ph} = 1/0.05$ ps$^{-1}$), the radiative efficiencies of SQ and SQMC can be approximated, within the limits of the rate equations model used, by $\Phi_{SQ} \approx \gamma_0/\gamma_{NR}$ and $\Phi_{SQMC} \approx \gamma_L/2\gamma_{NR}$, respectively, giving an estimation for the enhancement factor $G_e = \Phi_{SQMC}/\Phi_{SQ} \approx \gamma_L/2\gamma_0$. Assuming a value of $\gamma_L$=1/50 ps$^{-1}$, which is in agreement with previous theoretical works,[6, 17] and $\gamma_0$=1/2000 ps$^{-1}$,[43] we obtain an enhancement factor $G_e$≈20.[44]

## 3. Conclusion

We have fabricated organic microcavities by using highly reproducible thermal evaporation technique and demonstrated the enhanced emission intensity of a squaraine dye embedded in a low-Q cavity under strong coupling regime. These results cannot be explained by a modification of the vacuum field fluctuations, which become significant only in structures with smaller modal volumes and higher Q factors.[45-48] Moreover, our results are substantially different from the directional enhancement and spectral narrowing observed in devices working in the weak coupling regime,[30, 31] being the enhancement associated to a change in the emission dynamics, as shown in Figure 4, and occurring for all the accessible k-vectors. In addition, the decay rate perturbations of the dye due to the plasmonic environment have been also computed for the SQMC and SQ reference systems in the framework of the Discrete Dipole Approximation (DDA),[49] showing negligible changes under the experimental conditions.

The observed behavior is therefore associated with the formation of polariton modes which modify the energy levels of the system under strong coupling regime. The maximum enhancement occurs when the bottom of the LPB is at a lower energy than that of the dye emission peak (Figure 2), and with a comparable amount of exciton and photon components at k=0 (Figure S2 of the Supporting Information), supporting the interpretation of non-radiative, phonon-assisted scattering from an exciton reservoir into the bottom of the LPB. The kinetics are therefore expected to be more favorable at room temperature and depend on the LPB offset, according to what observed also in polymers and crystalline microcavities.[12, 25, 41, 42] Under these conditions, the absorption level, dressed with the photon component, is an efficient radiative channel which can contribute to reduce the non-radiative losses of the system. In general, the requirement of small Stokes shift and narrow linewidth for the observation of polariton emission seems to be unnecessary. On the contrary, the presence of a large Stokes shift could even be beneficial to increase the probability of radiative polariton emission with respect to the non-radiative relaxation of the molecule. High polariton population in the LPB could be thus achieved also in materials with large absorption coefficient but small emission efficiency.

## 4. Experimental Section

*Experimental Details*: The excitation power used in cw experiment is of less than 0.5 mW with an excitation density of less than 50 mW cm$^{-2}$. The pulsed experiments are performed with a frequency-doubled TiSa laser, pulsating at 80 MHz with a pulse length of 150 fs. The power used in the pulsed experiments are kept low enough to avoid degradation of the sample and where no changes in the dynamics are observed as a function of power. In the time-resolved experiments shown in the text, the excitation spot of the pulsed laser is of 0.5 mm$^2$, with a peak power per pulse of 10 kW cm$^{-2}$.

*PLQY*: Photoluminescence quantum yields ($\Phi$) were calculated from corrected emission spectra obtained by an Edinburgh FLS980 Spectrofluorimeter equipped with a barium sulphate coated integrating sphere (4 inches), a 450W Xe lamp as light source, and a R928 photomultiplier tube as signal detector, following the procedure described in Reference [32].

*Discrete Dipole Approximation*: Simulations are done by approximating the emitter with a classical oscillating dipole put in the middle of an active layer with the experimental refractive index of the squaraine blend and oscillating parallel to the metal interface with an energy of 1.8 eV. The Ag mirrors are 35 nm thick and large enough to minimize the effects of the target finiteness on the dipole dynamics. Total, radiative and nonradiative decay rates are calculated by varying the thickness of the film from 120 nm to 150 nm. As expected, by moving from the single to the double mirror the changes in the quantum yield do not reproduce the behavior of the PL intensity in Figure 2. The lifetime and quantum yield are not significantly affected by the cavity, which, in the weak coupling regime, only redistributes guided modes of the film into out-coupled modes.

**Supporting Information**
Supporting Information is available from the Wiley Online Library or from the author.

**Acknowledgements**
We are grateful to P. Michetti, G. La Rocca, S. K. Cohen and M. Bamba for fruitful and inspiring discussions. This work has been funded by the MIUR project Beyond Nano and the ERC project POLAFLOW (grant 308136).

Supporting Information

**Polariton Induced Enhanced Emission from an Organic Dye under Strong Coupling Regime**

*Dario Ballarini\*, Milena De Giorgi, Salvatore Gambino, Giovanni Lerario, Marco Mazzeo, Armando Genco, Gianluca Accorsi, Carlo Giansante, Silvia Colella, Stefania D'Agostino, Paolo Cazzato, Daniele Sanvitto\*, Giuseppe Gigli*

The cavity is composed by a thin film of co-evaporation of N,N'-bis(1-naphthyl)-N,N'-diphenyl-1,1'-biphenyl-4,4'-diamine (NPB) (molecule structure shown in **Figure S1**a) and 2,4-Bis[4-(N,N-diisobutylamino)-2,6-dihydroxyphenyl] (squaraine), in 90:10 volume ratio, sandwiched between two evaporated, 35 nm thick, semi-transparent silver mirrors. The HOMO and LUMO energy levels are -5.5 eV and -2.3 eV for the NPB and -5.3 eV and -3.6 eV for the squaraine, respectively (Figure S1b). In Figure S1c, the normalized absorption spectrum of the acceptor (red line) and emission spectrum of the donor (black line) are shown. The characteristic Foster distance is calculated to be $R_0 \approx 2$ nm, with the quantum yield of the donor $\approx 0.29$ and the molar absorption coefficient of the acceptor $\approx 300000$ $M^{-1}$ $cm^{-1}$. Transfer matrix calculations reproduce the formation of the UPB and LPB and are compared with the experimental transmittance measurements in Figure 1c of the text. The presence of the vibron at 2.05 eV and its effect on the polariton dispersion curves are instead shown in **Figure S2**a on a sample with cavity thickness of 135 nm. Background image is obtained from the measured transmission spectra as a function of the in-plane momentum, with the full dots indicating the point of maximum intensity. The white lines are the UPB and LPB as obtained by transfer matrix calculation considering one exciton at 1.85 eV with linewidth of 0.025 eV, while the red lines are the polariton dispersions obtained from transfer matrix by adding the second resonance at 2.05 eV (linewidth of 0.1 eV). The measured dispersion is better fitted when the second resonance is included, but the small oscillator strenght and the large damping parameter of the higher energy resonance limit to a small quantitative perturbation the difference between the two dispersion, confirming that the vibron is only weakly coupled with the cavity mode.

The excitonic and photonic components of polariton states are extracted by fitting the experimental results with the solutions:

$$U, L = \frac{\omega_0 + \omega_c}{2} - \frac{i}{2}(\gamma + \gamma_c) \pm \sqrt{\frac{(\omega_0 - \omega_c)^2}{4} + V^2 - \frac{(\gamma - \gamma_c)^2}{4} + \frac{i}{2}(\omega_0 - \omega_c)(\gamma_c - \gamma)} \qquad (S1)$$

The photonic and excitonic content of the lower polaritons at different k values are shown in Figure S2b for the case of the sample in Figure S2a, giving the minimum exciton fraction ≈0.4 at k=0. For film thicknesses in the range from 130 nm to 140 nm, the exciton component at the bottom of the LPB is between 50% (130 nm) and 30% (140 nm), while for thicker samples (see Figure 2b in the text), the more negative detuning corresponds to a larger photonic weight of the LPB, which helps, up to a certain extent, the emission of light, but inhibits the phonon-mediated scattering into the LPB. However, similarly to what happens in the case of inorganic microcavities, also the bottleneck effect reduces the scattering probability towards the bottom of the LPB, resulting in an overall decrease of the emission intensity at k=0 for thicknesses larger than 150 nm.

In the estimation of the enhancement, we note that the sample has not been optimized for the antinode factor, nor by choosing a bottom mirror of higher reflectivity, as discussed in Reference [31]. The intrinsic enhancement should be therefore at least two times higher than the experimentally observed one.

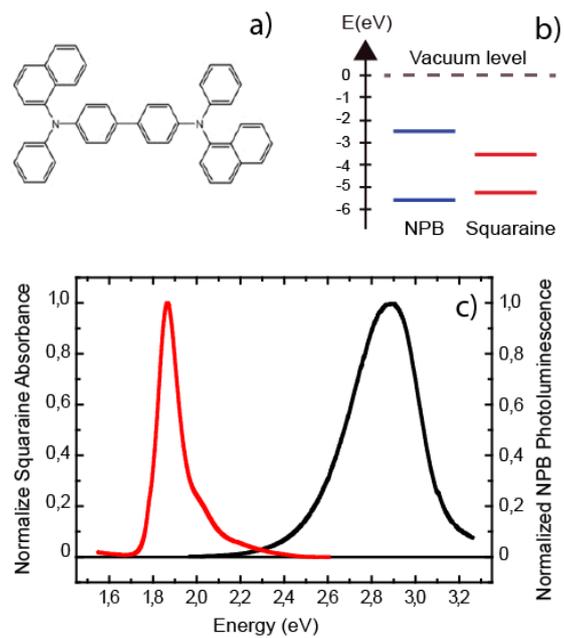

**Figure S1.**

a) Molecular structure of N,N'-bis(1-naphthyl)-N,N'-diphenyl-1,1'-biphenyl-4,4'-diamine (NPB). b) Representation of HOMO-LUMO energy levels for NPB (left) and squaraine (right) molecules. c) Normalized absorption spectrum of the acceptor (red line) and emission spectrum of the donor (black line) are shown.

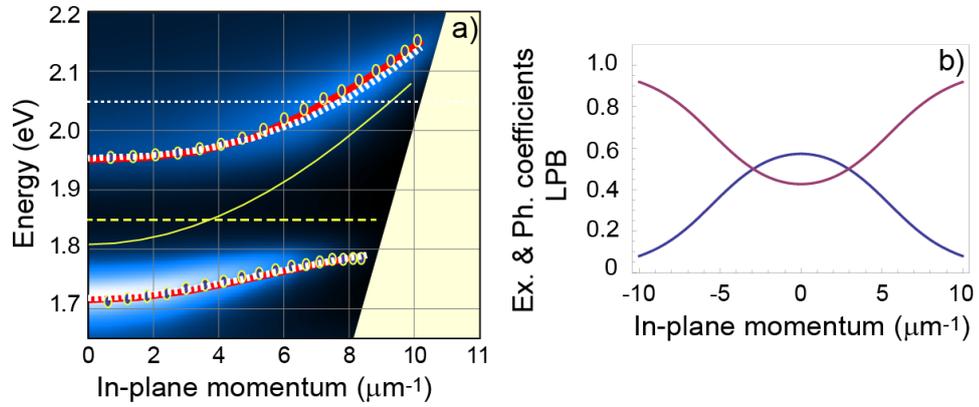

**Figure S2.**

a) The background image is obtained by the measured angular-resolved transmission spectra, plotted as a function of the in-plane momentum, with the dots indicating the position of the maxima along the dispersion. The UPB and LPB dispersion as obtained from transfer matrix calculations are shown as the dashed white lines for ω0=1.85 eV, γ0=0.025 eV and as solid red lines for ω0=1.85 eV, γ0=0.025 eV, ω1=2.05 eV and γ1=0.10 eV, respectively. b) Photonic (blue line) and excitonic (red line) content of the lower polaritons at different k values as obtained by fitting the dispersion shown in Figure S2a.

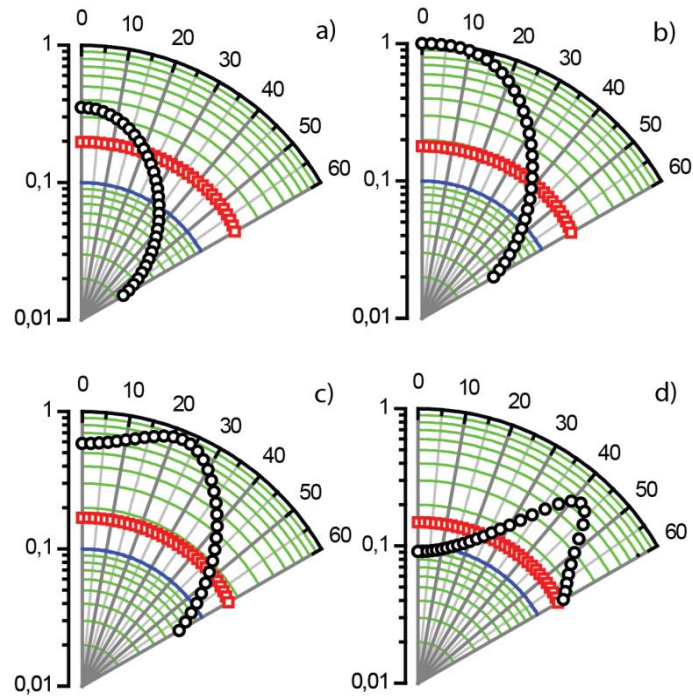

**Figure S3.**

Calculated directional enhancement and redistribution of the radiation pattern induced in the weak coupling regime by cavity structures of: a) 130 nm, b) 135 nm, c) 140 nm and d) 150 nm. Simulations are performed considering the experimentally measured mirrors parameters, and assuming a uniformly distributed, random-oriented emitting-dipoles in the cavity (only the real part n of the refractive index of the cavity has been used). Black and red points correspond to the angular intensities radiated out of the sample in the case of two mirrors and one mirror configuration, respectively. As expected, the cavity effect is strongly directional and induces at most an angular integrated (from 0° to 60°) enhancement of 2.5 times in the case of 140 nm thick cavity. Differently from Figure 3 in the text, here only the weak coupling regime is considered and the shown intensities (black and red points) correspond to the same wavelength of radiation (1.8 eV). As expected, the angular- and wavelength-integrated emission is similar for the one-mirror (1M) and two-mirrors (2M) configurations.

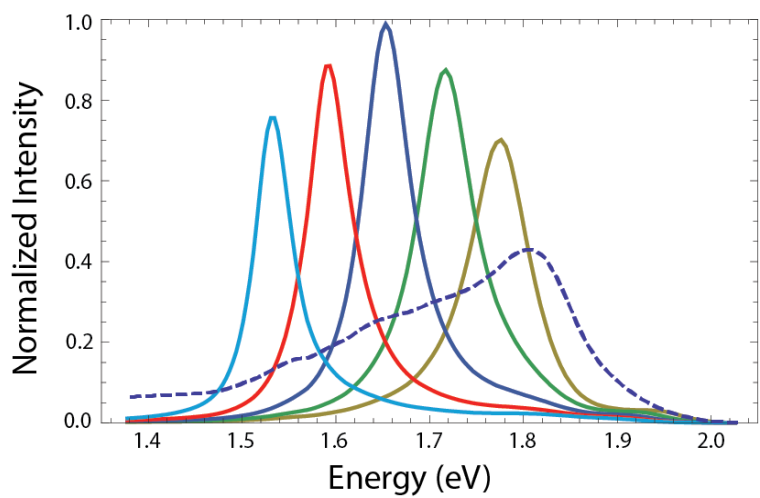

**Figure S4.**

Simulated dipole emission in the k=0 direction for one mirror (1M, dashed line) and two mirrors (2M, with light blue, red, blue, green and yellow corresponding to 160 nm, 150 nm, 140 nm, 130 nm and 120nm, respectively). In this simulations, uniformly distributed, random oriented emitting dipoles with the measured PL spectrum of SQ have been assumed. The complex refractive index of the film n,k used in the calculations has been obtained from the experimental absorbance by means of a Kramers-Kronig procedure. The changes in the dielectric constant due to the strong coupling are directly taken into account, showing the expected radiation of the SQ in the cavity through the polariton modes. While simulations can exactly reproduce the experimental transmittance spectra, the integrated intensity of simulated 2M emission are comparable with that of 1M for all the thicknesses considered. In addition, differently from Figure 2, the integrated intensity of 2M is only slightly affected (less than 25%) by changing the cavity thickness in the range between 120 nm and 160 nm. The difference with the experimental data shown in the text demonstrates the essential role played by scattering and relaxation dynamics in the enhancement of the emitted intensity.

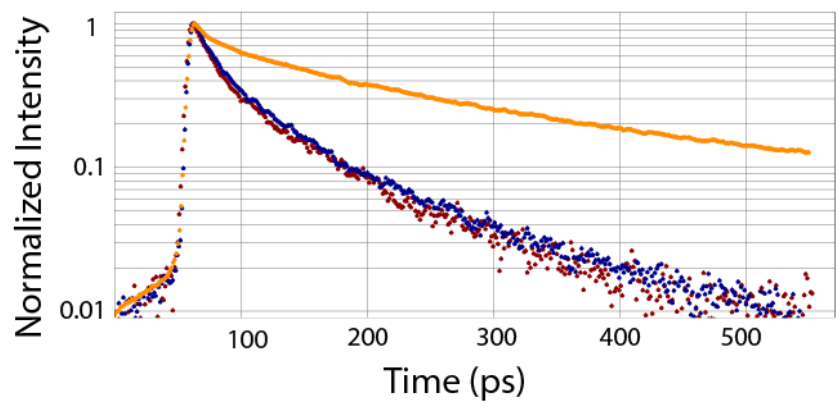

**Figure S5.**

TRPL detected at 2.8 eV (emission energy of NPB) in the case of the reference sample (red), the SQMC (blue) and neat film of NPB (orange). The decay is shorter when the squaraine dye is present in the blend, but exactly the same behaviour is observed in the case of SQ and SQMC samples.